\documentclass[aps, prd, twocolumn,
		nofootinbib,
		showpacs,showkeys,
		superscriptaddress]{revtex4-1}

\usepackage{epsfig,amssymb,bm,graphics,color}
\usepackage{epstopdf}
\usepackage{dcolumn}
\usepackage{amsmath}
\usepackage{bm}
\usepackage{slashed}
\usepackage{xcolor}
\usepackage[english]{babel}

\begin{document}

\title{Spectral caustics in laser assisted Breit-Wheeler process}

\author{T. Nousch}
\email{t.nousch@hzdr.de}
\affiliation{Helmholtz-Zentrum Dresden-Rossendorf, Institut f\"ur Strahlenphysik,
PF 510119, D-01314 Dresden, Germany}
\affiliation{Institut f\"ur Theoretische Physik, TU Dresden, D-01062 Dresden, Germany}

\author{D. Seipt}
\email{d.seipt@gsi.de}
\affiliation{Helmholtz-Institut Jena, Fr\"obelstieg 3, 07743 Jena, Germany}

\author{B.~K\"ampfer}
\affiliation{Helmholtz-Zentrum Dresden-Rossendorf, Institut f\"ur Strahlenphysik,
PF 510119, D-01314 Dresden, Germany}
\affiliation{Institut f\"ur Theoretische Physik, TU Dresden, D-01062 Dresden, Germany}

\author{A. I. Titov}
\affiliation{Bogoliubov Laboratory for Theoretical Physics, Joint Institute for Nuclear Research, 
RU - 141980 Dubna, Russia}

\date{\today}

\begin{abstract}
Electron-positron pair production by the Breit-Wheeler process embedded in a strong laser pulse is analyzed.
The transverse momentum spectrum displays prominent peaks
which are interpreted as caustics, the positions of which are accessible by the stationary phases.
Examples are given for the superposition of an XFEL beam with an optical high-intensity laser beam.
Such a configuration is available, e.g., at LCLS at present and at European XFEL in near future.
It requires a counter propagating probe photon beam with high energy 
which can be generated by synchronized inverse Compton backscattering.     

\end{abstract}


\maketitle

\section{Introduction \label{sect:in}}

Pair production processes in electromagnetic interactions are of permanent interest due to fundamental
aspects to be addressed up to technological relevance for material investigations. The basic process of
two-photon conversion into a pair of electron + positron, symbolically $X^\prime + X \to e^+ + e^-$
as $2 \to 2$ reaction of photons with four-momenta 
$k_{X^\prime, X} \sim (\omega_{X^\prime, X}, \mathbf k_{X^\prime, X}$) has been evaluated by 
Breit and Wheeler \cite{BW} within a framework which is called nowadays perturbative 
quantum electro dynamics (pQED). It is a t-channel process in lowest order pQED. The time-reversed
process is the famous annihilation, $ e^+ + e^- \to X + X^\prime$, widely used in medical applications
as positron emission tomography (better known under the acronym PET, cf.\ \cite{Parodi2008}) and 
material research \cite{Wagner,KrauseR}. In particle physics, the $\gamma$ conversion,
$X^\prime + X^* \to e^+ + e^-$ with $X^*$ referring to a (virtual) photon arising from an ambient medium,
e.g.\ from a static nuclear Coulomb field, is either a disturbing process calling for low-material budget
designs (e.g.\ \cite{HADES1}) or can be used for $\gamma$ detection purposes (cf.\ \cite{HADES2}
for an example).
There are many other elementary processes with emerging pairs which are accessible theoretically by
pQED, for instance such ones with $\mu^+ + \mu^-$ in the final state \cite{Titov},
or even with $\bar \nu + \nu$ \cite{Titovnu}.

Pair production is a threshold process, meaning that a certain minimum energy must be provided
in the entrance channel to have $e^+ + e^-$ with energy $> 2 m$ in the exit channel ($m$ is
the electron rest mass). This implies that the energies of the $X^\prime$ and $X$ photons must be
sufficiently large to overcome the threshold, i.e.\ 
$s_{X^\prime X} = (k_{X^\prime} + k_X)^2 
= 2 \omega_{X^\prime} \omega_X ( 1- \cos \theta_{X^\prime X}) >  4 m^2
\equiv s_\mathrm{thr}$,
with the relative angle  $\theta_{X^\prime X}$ of both beams is $\pi$ for head-on collisions. 
In the considered $2 \to 2$ scattering process, 
$s_{X^\prime X}$ equals the invariant mass $M^2=(p_e+p_p)^2$ of the produced electron-positron pair.

In case the initial center-of-mass energy is below the production threshold of the $2 \to 2$ process, 
$s_{X^\prime X} < s_\mathrm{thr}$, pairs can still be produced via multi-photon effects. 
This particularly interesting process has been investigated in the SLAC experiment E-144 \cite{Burke,Bamber},
where a high-energy photon (several GeV) was colliding with an intense optical laser pulse (L). 
While the $2 \to 2$ reaction was kinematically forbidden, 
the multi-photon channels $X+n L \to e^+ + e^-$ with $n>1$, had sufficient center-of-mass energy $s_{X,nL}=(k_X+nk_L)^2$ to overcome the pair production threshold. 
This process is called laser-induced multi-photon Breit-Wheeler pair production. 
In fact the high-energy photon was produced via Compton backscattering of laser light on 
$46.6$ GeV electrons in the same laser focal spot. (For a recent theoretical re-analysis see e.g.~Ref.~\cite{Keitel}.)
The multi-photon channels only have a considerable probability if the laser pulse is sufficiently intense.

The laser intensity parameter $a_0 = |e| E_L/m \omega_L$ (with $-\vert e \vert$ as the electron charge, 
and $E_L$ and $\omega_L$ refer to the field strength and frequency of the laser) 
delineates the non-relativistic domain, $a_0 < 1$, and the relativistic domain, where $a_0 > 1$ \cite{Piazzarewiev}. 
Moreover, $a_0$ quantifies the relevance of multi-photon effects; 
it is the inverse Keldysh adiabaticity parameter of the process. 
Another important parameter that classifies the pair production is the 
non-linear quantum parameter $\chi_\gamma = \frac 12 a_0 s_{X,1L} / s_\mathrm{thr}$ 
that combines $a_0$ and the kinematics of the process.
For $a_0 \lesssim 1$ and $\chi_\gamma \lesssim 1$ only a few multi-photon channels contribute, 
and the probability for the $n$th (open) channel behaves roughly as $W_n \sim a_0^{2n}$. 
For $a_0 \gg 1$ and $\chi_\gamma \lesssim 1$ (i.e.\ the $2 \to 2$ process 
is extremely deep below the threshold and huge amounts of laser photons are required) 
behaves semi-classically \cite{Meuren}.
The formation region of the pair becomes much shorter 
than the laser cycle, $\propto 1/a_0$, and the process takes place instantaneously 
as it were in a local constant crossed field. For $\chi_\gamma \ll 1$ 
the Breit-Wheeler pair production probability is exponentially suppressed in the semi-classical regime, 
$W\sim e^{-8/3\chi_\gamma}$ \cite{Reiss1962},
 with the same functional dependence on the electric field strength as Schwinger pair production 
\cite{Ritus,Sauter,Schwinger,Brezin,Popov}. For Schwinger pair production, the impact of an assisting high-frequency field has been studied, 
e.g.\ in \cite{Schuetzhold2008,Hebenstreit2014,Otto2015}.

The laser-induced multi-photon Breit-Wheeler process has been investigated exhaustively (see e.g. 
Refs.~ \cite{Ritus,Reiss1962,brown_1964}) for long-duration pulses of the laser beam. 
The process becomes markedly modified for ultra-short laser pulses: 
The temporal pulse structure, even in the plane-wave limit, gives a dominating impact specific for the pulse shape 
\cite{Nousch,Titov3,Titov31,krajewskaja2014,krajewskaja_1_2012}. In a finite pulse of the laser beam there are several interfering effects: finite bandwidth
(i.e.\ $\omega_L$ is the central frequency and higher and lower  frequencies contribute to the
power spectrum), multi-photon effects (i.e.\ the above mentioned higher harmonics) 
and the intensity-dependent threshold shifts \cite{kohlfurst,heinzl2010}.

Due to the small frequency of optical lasers, ${\omega_L = \mathcal {O}(1 {\rm \ eV})}$, the parameter $\chi_\gamma$ 
is very small unless the frequency of the colliding photon $X$ is very high -- on the order of several GeV. 
This makes the non-linear Breit-Wheeler pair production exceedingly small in pure optical laser-laser collisions 
unless both lasers have ultra-high intensities 
(of the order of the Sauter Schwinger field 
$\approx 10^{29} \, {\rm W} / {\rm cm}^2$)
\cite{Schwinger,Sauter, Ritus, brown_1964}. 
With the advent of x-ray free electron lasers (XFELs) 
that can provide photons with $\omega_X = \mathcal {O}(10{\rm \ keV})$ at high intensities, 
the gap to the threshold is diminished, but still fairly large, unless ${\omega_{X'} = \mathcal {O}(50{\rm \ MeV})}$. 
Therefore, one can ask whether the assistance of an ultra-high intensity laser beam $L$ enables pair production 
if $s_{X'X}$ is in the sub-threshold region. Clearly, also here, very strong non-linear effects 
due to an ultra-high intensity laser beam are required for enabling the this 
\textit{laser-assisted Breit-Wheeler pair production}. 
A related issue is the modification of the Breit-Wheeler process by an assisting laser beam above the threshold.

To attempt a description of this latter special process, we consider here the reaction
$X^\prime + (X + L) \to e^+ + e^-$, that is the laser assisted linear Breit-Wheeler process, where
$s_{X^\prime X} > s_\mathrm{thr}$ and $X$ is a weak field in the sense of $a_X \ll 1$; the probe photon
field is anyhow considered as weak, $a_{X^\prime} \ll 1$, i.e.\
only one photon from the field $X$ participates in a single pair production event.  
We have in mind the combination of an
XFEL  beam $X$ with a synchronized, co-propagating laser beam $L$ which may be strong. To be
specific, the intensity parameter of $X$ is less than $a_X = {\cal O}(10^{-2})$ according
to \cite{Ringwald2001}, and for the $L$ beam from a PW-class laser we let be $a_L = {\cal O} (1)$.
Note that $a_{X, L}$ depend on the size of the actual focal spots. Our considerations below
apply to the homogeneity region where a plane-wave approximation holds, but we include the
temporal pulse shape as an essential element. Considering the European XFEL beam, under construction
(and near to completion)  in Hamburg/DESY
\cite{XFEL}, in the HIBEF project \cite{HIBEF} with $\omega_X = 6$ keV, 
the counter-propagating beam $X^\prime$ must
have about $\omega_{X^\prime} = 60$ MeV  
(accessible, for instance, by suitable inverse Compton back-scattering of laser light off laser-accelerated electrons 
\cite{Jochmann2013,Powers2014,Leemans2014,Rykovanov2014,Sarri2014,Corde2013}) 
to allow for the linear Breit-Wheeler process. In the 
equal-momentum frame, $\mathbf k_{X^\prime} = - \mathbf k_X$, we have 
$\omega_{X^\prime} = \omega_X = 600$ keV and $s_{X^\prime X} / s_\mathrm{thr} = 1.38$.
For the assisting laser field we assume an UV laser frequency of 10 eV in the laboratory frame, i.e. $\omega_L = 1$ keV in the equal momentum frame.     
In this set-up, the pairs can not be produced by the $X^\prime - L$ collisions alone: This process
is extremely below the threshold and, thus, extremely suppressed since
$s_{X^\prime L} / s_\mathrm{thr} = 0.002$ and $\chi_\gamma = 0.001 a_0$.  
          
Our analysis is in many aspects parallel to 
\cite{Seiptx,Seiptcaustics}, where the laser assisted Compton
process is analyzed. This cross channel enjoys some remarkable features: The spectrum of Compton
scattered x-ray photons off an electron moving in an accelerated manner in an external laser pulse
displays, besides the well-known Compton line at fixed observation angle, a number of prominent
peaks, and the complicated spectral distribution exhibits distinct regions with changing patterns.
The striking finding in \cite{Seiptcaustics} is the interpretation of the prominent peaks as spectral caustics
related to merging stationary phase points. Accounting for quantum interference effects for the emission
from different locations of the quasi-classical electron motion in the laser field along a temporally changing
figure-8 trajectory, the gross features of the complicated spectrum become easily accessibly. Such an
interpretation is also in the spirit of \cite{Meuren}, where the spectrum of pairs produced in a strong
external field\footnote{
The interested reader is referred to  \cite{jansen_pair_2015,Ilderton2015,Lebed2011,Wu2014,Villalba-Chavez2013,Hatsagortsyan2011,Sokolov2010,DiPiazza2004}
for further work on pair production
in external fields within a QED framework.}
is explained as redistribution in phase space following the production process (which can
be approximated by a temporarily constant cross-field probability) and keeping interference effects.      
Such an interpretation, in turn, resembles effective models in strong-interaction processes, where the
amplitude is decomposed into an initial state interaction ($ISI$), followed by a hard production process
($HPP$), in turn  followed by a final state interaction ($FSI$), symbolically $ISI \times HPP \times FSI$
\cite{Kaptari1,Kaptari2,Titovbk}. Such a factorization is anchored in the Migdal-Watson theory (cf.\ \cite{MW}).

Despite of the similarities of the Compton and Breit-Wheeler processes related by crossing symmetry,
the different  phase spaces and attributed kinematic relations make them fairly different. This is the reason
for considering separately the analog of the spectral caustics in \cite{Seiptcaustics} in the
laser assisted Breit-Wheeler process. In the above spirit of the Migdal-Watson theory, the laser
assisted Compton scattering may be considered as based on the amplitude 
$ISI \times HPP_{\omega^\prime}$, while the laser assisted Breit-Wheeler process is
$HPP_p \times FSI$, with production amplitudes $HPP_{\omega^\prime, p}$ related by crossing symmetry.
$ISI$ and $FSI$ refer here to the motion of the charged particles in the laser field.

Our paper is organized as follows. In section II we present the QED basics for the calculation
of the laser assisted Breit-Wheeler process. Selected numerical results are discussed in section III
for a special kinematic situation to highlight the impact of the laser field.
Section IV summarizes.  

\section{The QED process}

In the Fury picture, the process $X^\prime + (X + L) \to e^+ + e^-$ is described by a one-vertex
diagram $X^\prime \to e^+_{X+L} + e^-_{X+L}$, where $e^\pm_{X+L}$ mean the Volkov solutions
of electrons and positrons in temporarily shaped fields $X + L$, both ones co-propagating and with
perpendicularly linear polarization. We consider head-on collision of the photons $X$ and $X'$. 
These assumptions are made for the sake of simplifications of the
subsequent evaluations. In addition, we linearize in the field $A_X$. This corresponds then to a
Furry-picture two-vertex $t$-channel diagram analog to the Breit-Wheeler process 
$X^\prime + X \to e^+_L + e^-_L$, where however the out-going electron (e) and positron (p) and the propagator
are laser dressed. 

The energy-momentum balance for laser-assisted pair production can be put into the form
($\mu$ is a Lorentz index)
\begin{equation}
k_{X^\prime}^\mu + k_X^\mu + \ell k_L^\mu = p_p^\mu + p_e^\mu \,,
\label{eq:emc}
\end{equation}
where $\ell$ represents an hitherto unspecified
momentum exchange between the assisting laser field $L$ and the produced pair. 
We define light-front coordinates, e.g.
$x^\pm = x^0 \pm x^3$ and $\mathbf{x}_\perp = \left (x^1,x^2\right)$
and analogously the light-front components of four-momenta.
They become handy because the laser four-momentum vectors only have one
non-vanishing light-front component $k_{L,X}^- = 2\omega_{L,X}$.
In particular, Eq.~\eqref{eq:emc} contains the three conservation equations in light-front coordinates: $k_{X'}^+ = p_p^+ + p_e^+$
and $\mathbf p_e^\perp = - \mathbf p_p^\perp$.
Moreover, the knowledge of all particle momenta allows to calculate $\ell$ via the fourth equation
\begin{equation}
\ell  =  \frac{1}{\eta}\left(\frac{p^-_{p}+p^-_{e}-k^-_{X^{\prime}}}{k^-_{X}}-1\right) \,,
\end{equation}
with the frequency ratio $\eta = \omega_L/\omega_X \ll 1$.
Note that the variable $\ell$ can be related to $M^2 = (1 + \eta \ell) s_{X^\prime X}$,
where $M$ is the invariant mass employed in di-electron spectroscopy, cf.\ \cite{HADES1,HADES2}.

It is convenient to parametrize the produced positron's phase space by the following three variables:
(i) the momentum exchange parameter $\ell$,
(ii) the azimuthal angle $\varphi$ with respect to the polarization direction of the assisting laser field and 
(iii) the shifted rapidity
\begin{equation}
z =  \frac{1}{2} \ln
\left(
\frac{p_{p}^{+}}{p_{p}^{-}}
\right)
+\frac12 \ln
\left(
\frac{\left(1 + \eta \ell\right) \omega_X}{\omega_{X^\prime}}
\right).
\label{eq:z}
\end{equation}
The case $z=0$ distinguishes the \textit{symmetric} situation where
the longitudinal laser momentum is equally shared between the
electron and the positron. In particular, in the equal momentum frame
each particle acquires the longitudinal momentum $p_\parallel = (p^++p^-)/2= -\ell \omega_L/2$.
Treating $(\ell,z,\varphi)$ as independent variables completely specifies the
four-momentum $p_p$ of the produced positron by using Eq.~\eqref{eq:z} and
\begin{align}
p_{\perp p}^2 = m^2 \left(  \frac{1+\eta \ell}{ \cosh^2 z } \frac{ s_{X'X} }{s_\mathrm{thr}} - 1\right) \, .
\label{eq:pperp}
\end{align}
Moreover, Eq.~\eqref{eq:emc}
allows to eliminate the dependence on the produced electron's momentum $p_e$.

The laser pulses $X+L$ are described by the four-vector potential
\begin{equation}
A^{\mu}  =  \frac{ma_L}{e}g_{L}\left(\eta\phi\right)\epsilon_{L}^{\mu}\cos\left(\eta\phi\right)
+\frac{m a_X}{e} g_{X}\left(\phi\right)\epsilon_{X}^{\mu}\cos\phi\ 
\end{equation}
with
the transverse polarization four-vectors $\epsilon_{L,X}^{\mu}$ obeying $k_{L,X}\cdot\epsilon_{L,X}=0$
(a dot indicates the scalar product of four-vectors) and $\epsilon_X\cdot \epsilon_L=0$,
and the pulse envelope functions 
\begin{eqnarray}
g_{L}\left(\phi\right) & = & \begin{cases}
\mbox{cos}^{2}\left(\frac{\pi\phi}{2\tau_{L}}\right) &, \, -\tau_L \le \phi \le \tau_L\\
0 &, \, {\rm otherwise} \qquad \quad,
\end{cases}\\
g_{X}\left(\phi\right) & = & \mbox{exp}\left(\frac{-\phi^{2}}{2\tau_{X}^{2}}\right)
\end{eqnarray}
with the dimensionless pulse lengths parameters $\tau_{L,X}$. 
The invariant phase is defined as $\phi=k_X \cdot x = \omega_Xx^+$.

For small intensities $a_X$ of the x-ray laser, the linearized differential cross section
reads

\begin{multline}
\frac{\mbox{d}^{3}\sigma}{\mbox{d}z\mbox{d} \ell \mbox{d}\varphi}  
=  \frac{\eta r_{0}^{2}}{4\pi 
k_X \cdot p_{e} \int_{-\infty}^{+\infty}  \mbox{d}\phi \, g_{X}^2 ( \phi ) }
\\ \times
\frac{\left(1-\mbox{tanh}z\right)}{\mbox{cosh}^{2}z}\,
\sum
\vert {\cal M} \vert^2 \,,
\label{Xsection}
\end{multline}
where the sum runs over the unobserved spin degrees of freedom of the produced pair as well as
the polarization states of the incident photon $X'$, and
with scattering amplitude 
\begin{eqnarray}
{\cal M} & = &  \mathcal{J}_{X}\mathcal{A}_{0}-{\alpha_X}\sum_{k=0,1,2}\mathcal{J}_{k}\mathcal{A}_{k},\\
\alpha_X & = & \frac12 m \left( \frac{p_{e} \cdot \epsilon_{X}}{k_{X} \cdot p_{e}}
-\frac{p_{p} \cdot \epsilon_{X}}{k_{X} \cdot p_{p}}\right) 
\end{eqnarray}
and classical electron radius $r_0 = \alpha_\mathrm{QED} / m$, with the fine structure constant ${\alpha_\mathrm{QED} \simeq 1/137}$.
The spin and polarization dependence of the scattering amplitude
is encoded in the Dirac current structures
\begin{eqnarray}
\mathcal{J}_{0} & = & \bar{u}(p_e) \slashed\epsilon_{X^\prime} v(p_p), \label{J0}\\
\mathcal{J}_{1} & = & \bar{u}(p_e)a_L \left(d_p  \slashed\epsilon_L \slashed k_X \slashed\epsilon_{X^\prime}
-d_e \slashed\epsilon_{X^\prime}\slashed k_X \slashed\epsilon_{L}\right) v(p_p), \label{J1} \\
\mathcal{J}_{2} & = & 2d_e d_p  a_{L}^{2}\left( k_X \cdot \epsilon_{X^\prime} \right)
\bar{u}(p_e) \slashed k_{X} v(p_p) \label{J2}\\
\mathcal{J}_{X} & = & \bar{u}(p_e) 
\left(d_p \slashed\epsilon_{X} \slashed k_X \slashed\epsilon_{X^\prime} - 
d_e \slashed\epsilon_{X^\prime} \slashed k_X \slashed\epsilon_{X} \right) v(p_p), \label{JX} 
\end{eqnarray}
with the polarization four-vector of the probe photon $\epsilon_{X^\prime}$ fulfilling 
${k_{X^\prime}\cdot \epsilon_{X^\prime}=0}$. In addition, we defined 
${d_p=m/(2 k_X \cdot p_p)}$ and
${d_e = m/(2 k_X \cdot p_e)}$.
We employ the standard Dirac spinor wave functions $u, v$ and their adjoints
$\bar u, \bar v$ for electrons and positrons, respectively. 
Moreover, Feynman's slash notation is used. 
Note that the $\mathcal{J}_{k}$ defined in Eqs.~\eqref{J0}--\eqref{J2} are just complex numbers, albeit depending on momenta, polarizations and spins.

The dynamics of the laser-assisted pair production process is described by the integrals
\begin{equation} \label{Aj}
\mathcal{A}_j  =  \int_{-\infty}^{+\infty} d\phi
\left[ \cos (\eta\phi ) \, g_L (\eta\phi ) \right]^j  
g_X ( \phi) \, \mbox{e}^{i H} 
\end{equation}
for $j = 0, 1, 2$ (on l.h.s.\ a label, while on r.h.s.\ a power).
The phase $H$ of these integrals can be expressed as
\begin{equation}
H  = 
\int_{\eta\phi_{0}}^{\eta\phi}
d \phi^\prime 
\left( \ell+\frac{\alpha}{\eta}g_{L}\left(\phi^{\prime}\right)\cos\left(\phi^{\prime}\right) + \frac{\beta}{\eta}g_{L}^{2}\left(\phi^{\prime}\right)\cos^{2}
\left(\phi^{\prime}\right) \right) 
\label{eq:PhaseHL}
\end{equation}
up to an irrelevant arbitrary integration constant $\phi_0$ 
and using the abbreviations
\begin{align}
\alpha & =  m a_L
\left(\frac{p_e \cdot \epsilon_{L}}{k_X \cdot p_e} - \frac{p_p \cdot \epsilon_{L}}{k_X \cdot p_p}\right),
\label{def_alpha} \\
\beta & =  \frac{\left(ma_L \right)^{2}}{2}
\left(\frac{1}{k_X \cdot p_e}+\frac{1}{k_X \cdot p_p}\right) .
\label{def_beta}
\end{align}
Note that the phase $H$ can be rewritten directly in terms of the classical trajectory of the generated positron
moving in the assisting laser field,
projected onto the four-momentum vector $k_{X^\prime}$ of the probe photon.
This suggests the interpretation of the production by a plain Breit-Wheeler process
${X^\prime + X \to e^+ + e^-}$ as the hard production process (\textit{HPP})
with a subsequent redistribution of the positrons
in phase space due to the action of the laser field (referred to as final state interaction (\textit{FSI})).
By integrating over $\phi$ in Eq.~\eqref{Aj} one coherently adds the production amplitudes from all
``instants'' (expressed by the laser phase) which---after the redistribution due to \textit{FSI}---contribute 
to the yield of positrons at the chosen final phase-space point ($\ell,z,\varphi$).

The stationary phase condition $d H / d \phi = 0$ reads, by means of (\ref{eq:PhaseHL}),
\begin{equation}
0 = \ell+\frac{\alpha}{\eta}
g_L ( \phi) \cos ( \phi )+\frac{\beta}{\eta} g_L^2 (\phi ) \cos^{2} (\phi ),
\label{eq:stationarity}
\end{equation}
representing an approximation w.r.t.\ the highly oscillating phase factor
$\exp (iH)$ in (\ref{Aj}).
The stationarity condition \eqref{eq:stationarity} furnishes a relation between the instant $\phi$
the pair is produced and the momentum exchange $\ell$.
In order to solve \eqref{eq:stationarity} for $\ell(\phi)$ we first need to work out how
the coefficients $\alpha$
and $\beta$ depend on $\ell$.
 Here and in the following we restrict our investigation
to those positrons that are detected in the polarization direction of the assisting laser,
characterized by $\varphi=\pi$.
By eliminating the electron momentum $p_e$ in Eqs.~\eqref{def_alpha} and \eqref{def_beta} with help of Eq.~\eqref{eq:emc} and by rewriting the scalar products in terms of the independent variables
$(\ell,z,\varphi)$ we find
${\alpha = -2\beta \sqrt{ (1 +\eta \ell) /\beta - a_L^{-2}} }$
and
${\beta = a_L^2 s_\mathrm{thr} \cosh^2 (z) / s_{X^\prime X}}$.
Using these expression in Eq.~\eqref{eq:PhaseHL} we obtain a quadratic equation for $\ell (\phi)$, 
in contrast to the laser-assisted Compton
scattering of x-rays studied in \cite{Seiptcaustics}, 
with the two apparent solutions
\begin{equation}
\ell_{\pm} (\phi) = 
\frac{\beta}{\eta}
g_ L (\phi) \cos (\phi ) \left[g_L (\phi ) \cos (\phi ) \pm 2 \sqrt{\frac{1}{\beta}-\frac{1}{a_{L}^{2}}}\right].
\label{ell_phi}
\end{equation}
One has to check, however, for which values of $\phi$ the $\ell_\pm(\phi)$ represent true solutions of the
initial Eq.~\eqref{eq:stationarity}.
These solutions for $\ell_\pm(\phi)$, which follow from the stationary phase condition, determine
the amount of laser momentum that is transferred to the positron after its production at the instant $\phi$, and finally
arriving at the phase-space point $(\ell,z,\varphi)$.
That means, positrons at some fixed $\ell$ in phase space are produced only at a few certain instants.

\section{Numerical results and interpretation as spectral caustics}

In Figs.~1 -- 3, upper panels, we show numerical examples of the differential spectra 
$d^3 \sigma /d \ell dz d \varphi$ for $z = 0$ and $\varphi = \pi$. 
The chosen 
$\sqrt{s_{X^\prime X}} = 1.2$ MeV is clearly above the threshold, $s_{X'X}/s_\mathrm{thr}=1.38$, and the Breit-Wheeler peak at $\ell = 0$ corresponding to\footnote{Since we consider here exclusively the positron out-states, the label "$p$" is dropped.}
$p_{\perp} = \frac12 \sqrt{s_{X^\prime X} -s_\mathrm{thr} }$
is visible as pronounced structure.

According to the semi-classical interpretation, and the \textit{HPP}+\textit{FSI} scheme,
\textit{all} positrons are generated at $\ell=0$ via the ``bare'' Breit-Wheeler process.
The assisting laser field acting on these positrons shifts them in phase space due to
the exchange of laser momentum and they end up at $\ell\neq0$.
Consequently, the spectrum of the positrons that is observed after the interaction with the laser
is spread out essentially between the cut-off values
$\ell_\mathrm{min} \leq \ell \leq \ell_\mathrm{max}$
(or equivalently $p_\perp^\mathrm{min} \leq p_\perp \leq p_\perp^\mathrm{max}$).
Therefore, only for $\tau_X > \tau_L$ the ``bare'' Breit-Wheeler peak at $\ell=0$ is clearly visible, when
those positrons which are created before/after the laser impact remain at their place of birth in phase space.
For smaller values of $\tau_X$, the Breit-Wheeler peak vanishes since all positrons are shifted
away upon the subsequent laser action. 
The cut-off values can be determined from the minimum and maximum values $\ell_\pm$ in Eq.~\eqref{ell_phi}. They read
$\ell_\mathrm{max} = \ell^{(+)}$
and
$\ell_\mathrm{min} = \min 
\left( \ell^{(-)} , \ell_\mathrm{kin} \right)
$
with ${\ell^{(\pm)} = s_\mathrm{thr} a_L / (\eta s_{X^\prime X} ) 
\left( a_L \pm 2 \sqrt{s_{X^\prime X} / s_\mathrm{thr} - 1} \right)}$.
The lower cut-off is influenced by the fact that
the positron can lose at most its kinetic energy due to the laser action,
it needs to retain at least its rest energy.
That means $(1+\eta \ell ) s_{X'X} = M^2 \geq s_\mathrm{thr}$, yielding
the kinematic cut-off $\ell > \ell_\mathrm{kin} =  (s_\mathrm{thr}/s_{X'X} - 1 )/\eta$.
(The corresponding cut-off values for $p_\perp$ follow from Eq.~\eqref{eq:pperp}.)
Beyond the plateau region spanned by these cut-off values the spectra are going exponentially fast to zero.

The influence of the laser field intensity $a_L$ is evident upon comparing Figs.~1 and 2: 
The minimum and maximum
values of $p_\perp$, respectively $\ell$, do strongly differ. A few, albeit not all, strong peaks can be
attributed to caustics in the spirit of \cite{Seiptcaustics}: These are the loci of merging branches
of stationary phase points (see lower panels) determined by diverging $d \phi(\ell) / d \ell$
(see vertical lines in lower and
upper panels). Due to interference effects the quasi-classically determined caustic positions    
do not necessarily exactly coincide with the peak positions.
The gray bands depict the estimated 
widths of the caustic zone by $\Delta \ell = (a_L / \eta)^{2/3}$, 
following from the universality of the caustic's properties \cite{Kravtsov1983,Seiptcaustics}.

The shape of the differential spectra in the region around the spectral caustics resembles
indeed the caustics known from diffraction: They show the typical behavior of an 
Airy function describing the intensity distribution of light close to an optical caustic, e.g. that of the rainbow \cite{Airy1838}. 
This behavior is most pronounced at the upper cut-off values because only the caustic contributes there. 
For all the other peaks, the caustic contributions are accompanied by non-caustic contributions from the 
other branches of $\phi(\ell)$. This is particularly evident in Fig.~2. 
Moreover, the highly oscillatory behavior of the spectra can be explained as the interference from 
the contributions from the multiple stationary points.

The impact of the laser pulse length $\tau_L$ is obvious in comparing Figs.~2 and 3: The patterns of $\phi(\ell)$ are
different (see lower panels) and, correspondingly, the spectra too (see upper panels). The
shorter pulse implies fewer caustics with clearer correspondence to the prominent peaks in the transverse
momentum spectra.
At smaller values of $a_L$, the estimated widths of the caustics become too large and overlapping
thus not supporting the caustical interpretation of the spectra. At larger values of $a_L$
(e.g.\ $a_L \ge 3$) additional spectral modulation
effects caused by the beating of the $\pm$ branches in (\ref{ell_phi}) deserve separate investigations.  

\begin{figure}
\includegraphics[width=0.48\textwidth]{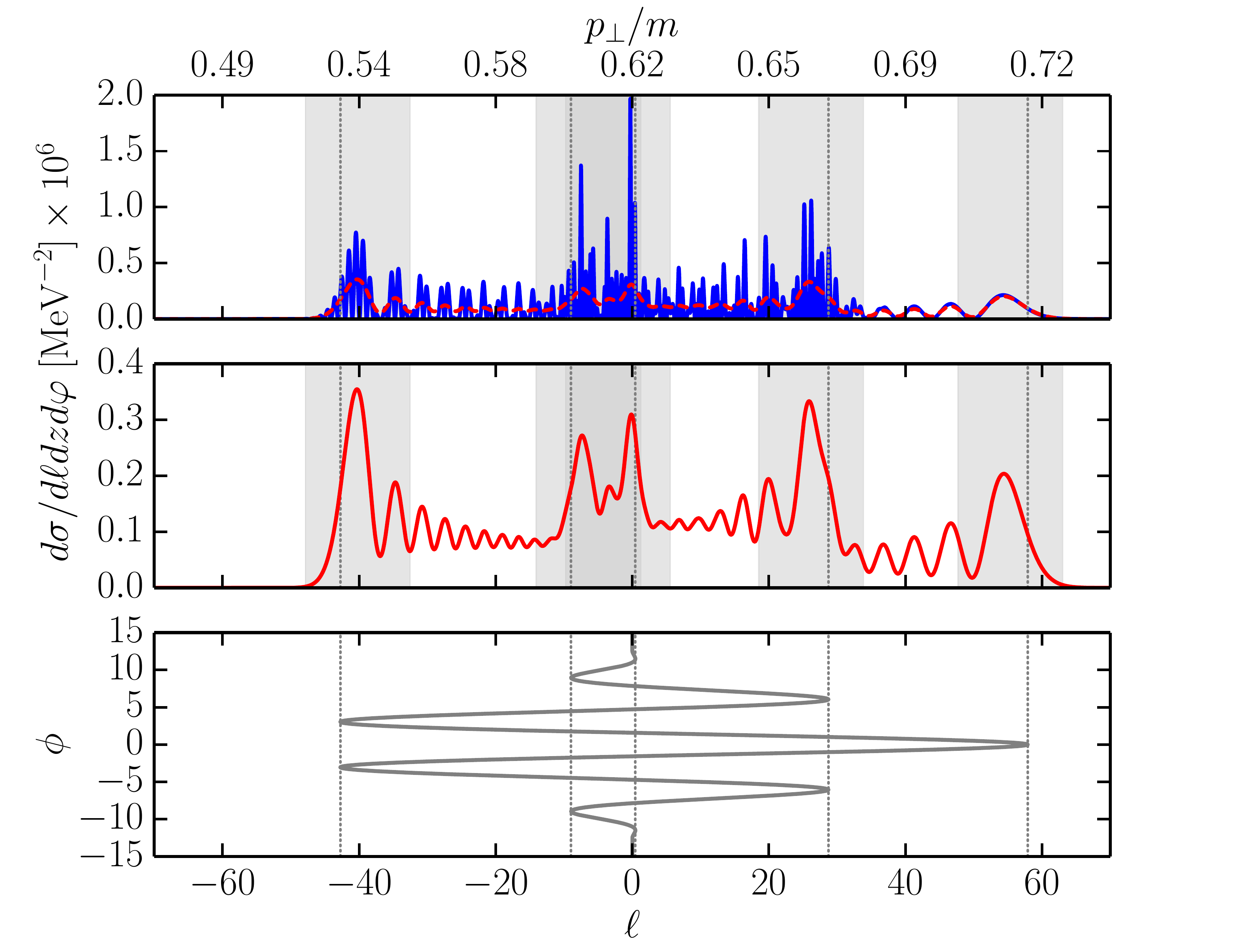}
\caption{
Spectra for the laser assisted Breit-Wheeler process with the parameters
mentioned in the Introduction which translate into 
$\sqrt{s_{X^\prime X}} = 1.2$ MeV,
$\eta = 1/600$,
${a_X = 10^{-5}}$, $\tau_X = 2 \tau /(\pi \eta)$, 
$a_L = 0.1$, and $\tau_L =4 \pi$
in the field (2). 
Upper panel: $d \sigma /d \ell d z d \varphi$ at $z = 0$ and $\varphi = \pi$
as a function of $\ell$ (lower axis; the corresponding values of $p_\perp$ are given
at the upper axis). The calculated spectrum according to (\ref{Xsection})
(blue, with 20,000 meshes) is smoothed by a Gaussian window function with width $\delta \ell = 0.8$
to get the red curve. 
Middle panel: smoothed spectrum separately.
Lower panel: phase $\phi$ as a function of $\ell$ from Eq.~(\ref{ell_phi})
(only the ``$+$'' solution applies here).
The vertical dotted lines depict the
positions of diverging $d \phi / d \ell$, where two branches of $\phi (\ell)$ merge.
The gray bands depict the estimated widths of caustic regions.   
\label{fig:1}}
\end{figure}

\begin{figure}
\includegraphics[width=0.48\textwidth]{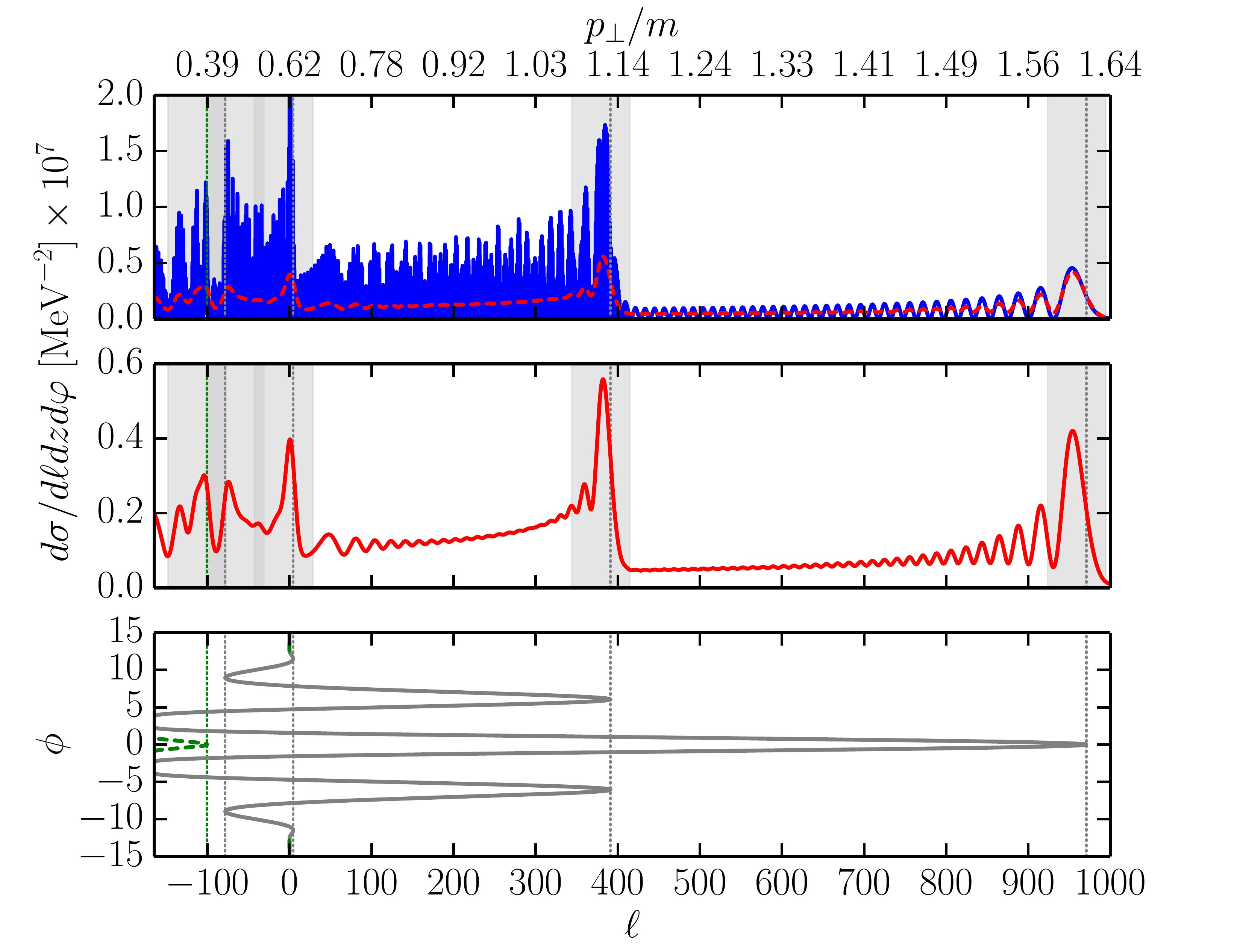}
\caption{As Fig.~1 but for $a_L = 1$ and $\delta \ell =5$.
In bottom panel, in black the ``$+$''
solution of (\ref{ell_phi}), while the green dashed curve is for the ``$-$'' solution.
The left boundary of the figure corresponds to the kinematic cut-off 
$\ell_\mathrm{kin}$, or equivalently $p_\perp=0$.
\label{fig:2}}
\end{figure}

\begin{figure}
\includegraphics[width=0.48\textwidth]{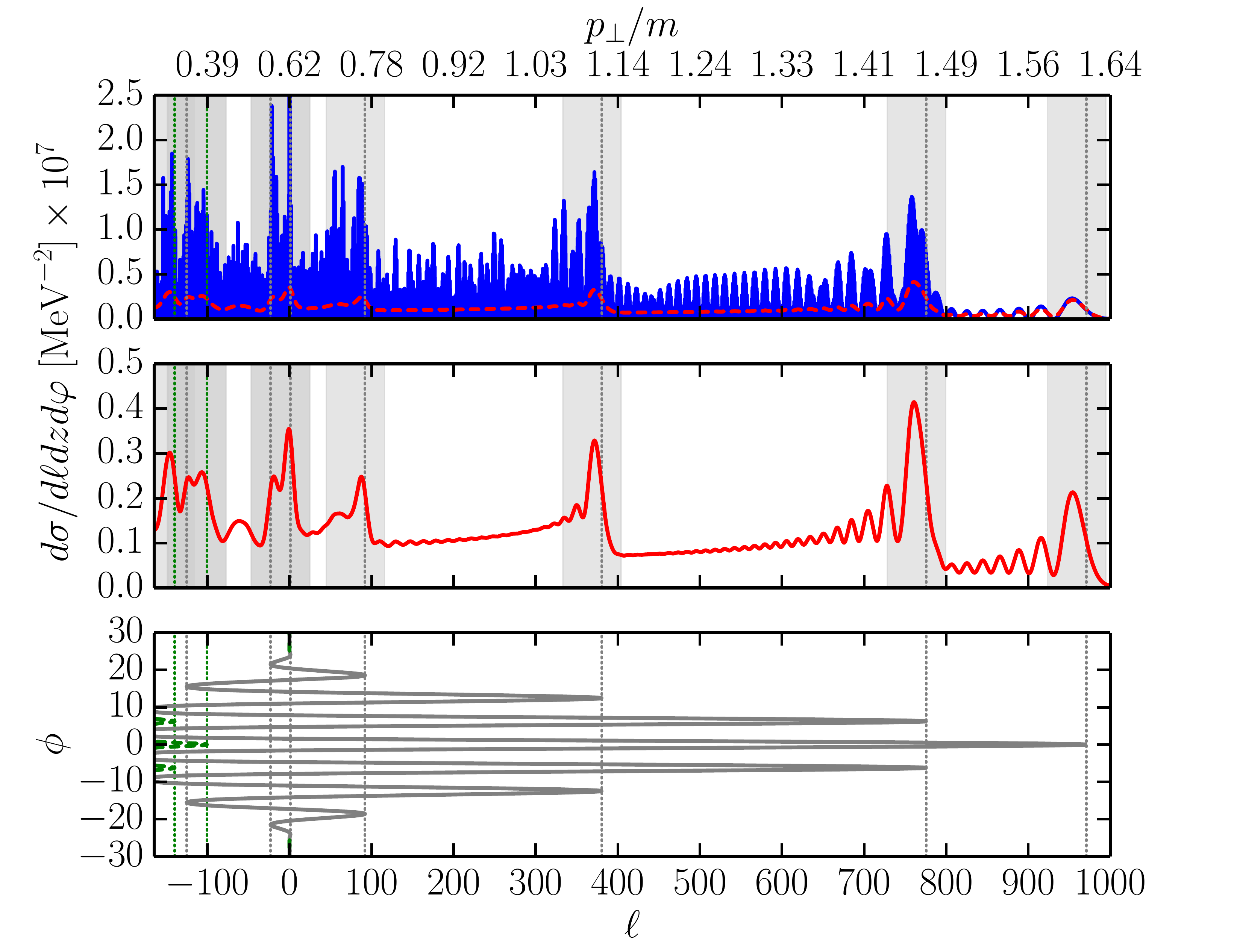}
\caption{As Fig.~2 but for the longer laser pulse duration $\tau= 8 \pi$, resulting in a larger number
of caustic peaks as compared to Fig.~2.
\label{fig:3}}
\end{figure}

\section{Summary}

In summary we show that the differential spectra, most noticeably the transverse momentum distributions
at fixed rapidity (more precisely, at $z = 0$ and fixed azimuthal angle of the positron) in laser assisted
Breit-Wheeler pair production is strikingly modified by details of the lase pulse shape. In the spirit of the
Migdal-Watson theory one may attribute this phenomenon to a final state interaction of the once
produced charged particles in the laser field. In other words, the quasi-classical motion with account of
interference effects offers a key to the gross features of the spectra. On the one hand, the manifestation
of the trajectories is not so surprising since the phase of the employed Volkov solutions for the
$e^\pm$ wave functions encodes the classical Hamilton-Jacobi action. On the other hand, the convolution
with other kinematic quantities of the squared matrix element is not so strong to destruct this
trajectory information. The interpretation of the series of distinct peaks as spectral caustics, analog to
laser assisted Compton scattering of x-rays, is semi-quantitative  since obviously severe interference effects of the
quantum mechanical propagation from certain phase points are, in general, responsible for the highly
non-trivial final momentum distribution.

Finally, we speculate that the trident process, i.e.\ the seeded pair production in a virtual Compton
process, may exhibit similar momentum signatures which could be also interpreted as spectral caustics.
Corresponding experiments are possible with the set-ups planned by the HIBEF collaboration.          
  
\acknowledgments
 The authors acknowledge fruitful discussions with
 R.~Sauerbrey and T.~E.~Cowan within the HIBEF project
and A. Di Piazza, C. H. Keitel, H. R. Reiss, V. G. Serbo, R. Sch\"utzhold, G. Dunne, D. Blaschke, C. M\"uller, R. Alkofer,
T. Heinzl, S. Frizsche, and A. Surzhykov on elementary strong-field QED processes of contemporary interest.

\bibliography{TNousch_biblio}

\end{document}